\input psfig.sty
\def\ho{{$^1$H}}
\def\het{{$^3$He}}
\def\hef{{$^4$He}}
\def\ctw{{$^{12}$C}}
\def\cth{{$^{13}$C}}
\def\nf{{$^{14}$N}}

\def\thn{{\thinspace}}
\def\scr{\scriptstyle}
\def\ref{\par \noindent \hangindent=3pc \hangafter=1}

\def\Msun{\hbox{$\thn M_{\odot}$}}

\def\={\thn\thn=\thn\thn}

\def\tgs{{\thn \rlap{\raise 0.5ex\hbox{$\scr  {>}$}}{\lower 0.3ex\hbox{$\scr  {\sim}$}} \thn }}
\def\tls{{\thn \rlap{\raise 0.5ex\hbox{$\scr  {<}$}}{\lower 0.3ex\hbox{$\scr  {\sim}$}} \thn }}
\def\tll{{\raise 0.3ex\hbox{$\scr  {\thn \ll \thn }$}}}
\def\tgg{{\raise 0.3ex\hbox{$\scr  {\thn \gg \thn }$}}}
\def\tle{{\raise 0.3ex\hbox{$\scr  {\thn \le \thn }$}}}
\def\tge{{\raise 0.3ex\hbox{$\scr  {\thn \ge \thn }$}}}
\def\tl{{\raise 0.3ex\hbox{$\scr  {\thn < \thn }$}}}
\def\tg{{\raise 0.3ex\hbox{$\scr  {\thn > \thn }$}}}
\def\ts{{\raise 0.3ex\hbox{$\scr  {\thn \sim \thn }$}}}

\def\do{{\tt "}}
\def\tp{{\raise 0.3ex\hbox{\small +}}}

\def\dm{\Delta\mu}
\def\tmix{t_{\rm mix}}
\def\tburn{t_{\rm burn}}
\def\dt{\Delta T}

\documentclass{iaus}
\usepackage{graphicx}
\title[$^3$He Destruction] 
{The Destruction of $^3$He by Rayleigh-Taylor Instability on the First Giant Branch}

\author[Eggleton, Dearborn \& Lattanzio]   
{Peter P. Eggleton$^{1,3}$, David S. P. Dearborn$^{2,3}$ \break \and John C. Lattanzio$^4$}

\affiliation{$^1$Institute of Geophysics and Planetary Physics\\
$^2$Physics and Allied Technologies Division\\
$^3$Lawrence Livermore National Laboratory, 7000 East Ave, \\
Livermore, CA94551, USA \break email: ppe@igpp.ucllnl.org, dearborn2@llnl.gov\\[\affilskip]
$^2$Centre for Stellar and Planetary Astrophysics,Monash University,
Australia
\break email: john.lattanzio@sci.monash.edu.au}

\pubyear{2007}
\volume{239}  
\pagerange{1--8}
\date{?? and in revised form ??}
\jname{Proceedings Title IAU Symposium}
\editors{F. Kupka \& I. W. Roxburgh, eds.}
\begin{document}

\maketitle

\begin{abstract}
Low-mass stars, $\ts 1-2$ solar masses, near the Main Sequence are efficient 
at producing \het, which they mix into the convective envelope on the giant 
branch and distribute into the Galaxy by way of envelope loss. This process 
is so efficient that it is difficult to reconcile the observed cosmic 
abundance of \het\ with the predictions of Big Bang nucleosynthesis. In this 
paper we find, by modeling a red giant with a fully three-dimensional 
hydrodynamic code and a full nucleosynthetic network, that mixing arises in 
the supposedly stable and radiative zone between the hydrogen-burning shell 
and the base of the convective envelope. This mixing is due to Rayleigh-Taylor 
instability within a zone just above the hydrogen-burning shell. In 
this zone the burning of the \het\ left behind by the retreating convective 
envelope is predominantly by
the reaction \het\ +\ \het\ $\to$ \hef\ +\ \ho\ +\ \ho, a reaction which,
untypically for stellar nuclear reactions, {\it lowers} the mean molecular 
weight, leading to a local minimum. This local minimum 
leads to Rayleigh-Taylor instability, and turbulent motion is generated which 
will continue ultimately up into the normal convective envelope. Consequently 
material from the envelope is dragged down sufficiently
close to the burning shell that the \het\ in it is progressively destroyed.
Thus we are able to remove the threat that \het\ production in low-mass stars 
poses to the Big Bang nucleosynthesis of \het.
\par Some slow mixing mechanism has long been suspected, that connects the
convective envelope of a red giant to the burning shell. It appears to be
necessary to account for progressive changes in the \ctw/\cth\ and \nf/\ctw\ 
ratios on the First Giant Branch. We suggest that these phenomena are also
due to the Rayleigh-Taylor-unstable character of the \het-burning region.
\end{abstract}

\firstsection 
\section{Introduction}

\par  Stellar evolution has long shown rather clearly that in the 
Main-Sequence (MS) region stars burn hydrogen in their cores by a 
combination of the pp chain
and the CNO tri-cycle. The former is the more important in low-mass stars,
$\tls 1.5\Msun$, and the latter in more massive stars. In the low-mass
stars much $^3$He is produced in a region outside the main H-burning
core, and because the convective core is small or absent in such stars
this $^3$He survives and is mixed (Iben 1967) into the Surface Convection 
Zone (SCZ) as the star ascends the First Giant Branch (FGB). The initial 
abundance of $^3$He may be increased above its primordial value, taken to be
$2\times 10^{-4}$, by a factor of nearly 10.
\par Fig. 1a illustrates the distribution of\ \het\ (and other isotopes) in a 
$0.8\Msun$ Pop II star towards the end of its MS life.  The Figure is
the result of a 1D (i.e. spherically symmetric) calculation. \het\ is 
enriched above its initial value (the same as its surface value since this star 
has only a slight convective envelope) in a broad peak extending over nearly
half the mass of the star. The peak abundance is a factor of $\ts 30$ larger
than the initial value. 

\vskip 0.1truein
{\psfig{figure=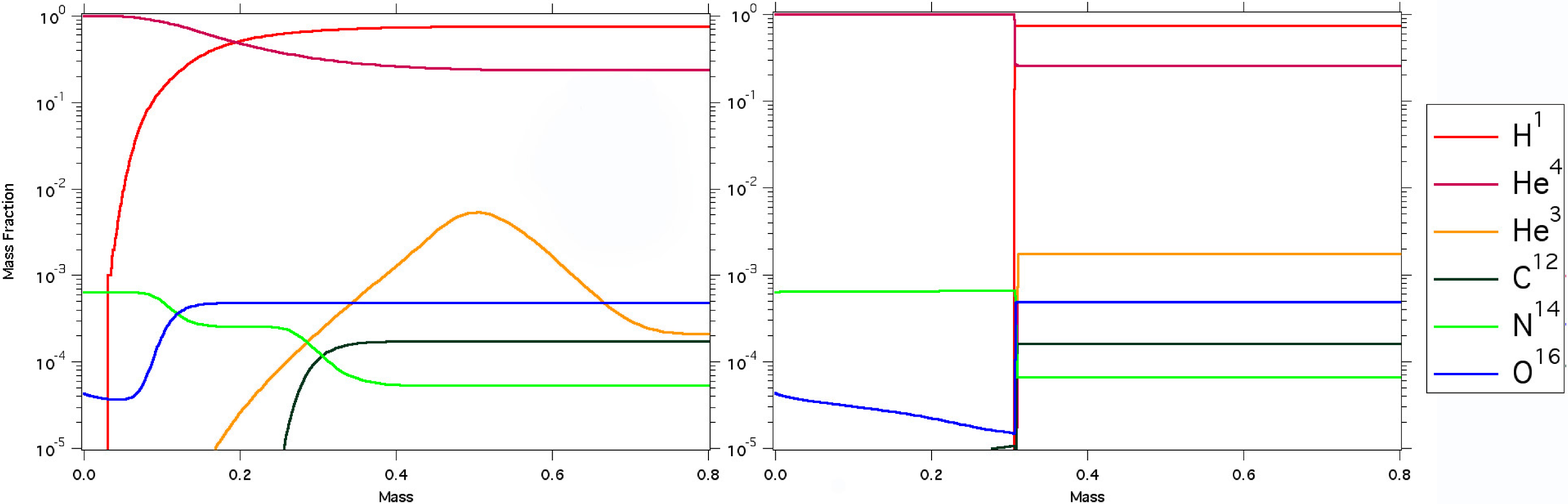,height=2.1in,bbllx=0pt,bblly=0pt,bburx=1950pt,bbury=750pt,clip=}}
\vskip 0.2truein
{\small {\bf Figure 1}.  (a) Profiles of the abundances of certain nuclei in a 
star which has evolved to roughly the end of the MS. Although we only model a 
Pop II star here, the enrichment of \het\  is just as considerable for 
relatively metal-rich Pop I stars like the Sun. \het\ shows a major peak
where the abundance reaches $\ts 30$ times the initial (surface) abundance.
(b) The same star later, when the SCZ reaches its maximum inward extent.
The \het\ peak has been homogenized, to a factor of 9 larger than its initial
value. The inert, H-depleted core is about $0.3\Msun$.}
\vskip 0.1truein

\par In later evolution, a large SCZ develops which
mixes and homogenises the outer $\ts 0.5\Msun$ (Fig. 1b). The surface \het\ 
abundance is raised from the initial $2.10^{-4}$ to $\ts 1.8.10^{-3}$, i.e. by 
a factor of $\ts 9$. Later still, as the star climbs the FGB, the
SCZ is diminished by (a) nuclear burning below its base, in a zone
that marches outwards, and (b) stellar-wind mass loss from its surface.
The evidence for the latter is that the next long-lived stage after the FGB
is the Horizontal Branch (HB), and HB stars appear to have masses that are
typically $0.5 - 0.6\Msun$, substantially less than the masses of stars
capable of evolving to the FGB in less than a Hubble time (Faulkner 1966, 1972). 
Process (b) leads to enrichment of the interstellar medium (ISM) in \het\ 
(Steigman et al. 1985, Dearborn et al. 1986, 1996).
\par Only relatively low-mass stars contribute to this enrichment, because in
massive stars the convective core becomes a large enough fraction of the total 
stellar mass that it mixes the \het\  peak to the center, where the \het\  is
burnt to \hef. Roughly, we expect that stars in the mass range $0.8 - 2\Msun$
are the ones that contribute to \het\ enrichment; but this is a substantial
majority, by combined mass as well as by number, of all stars capable of
substantial evolution in the Galaxy's lifetime. Yet the ISM's abundance of \het, 
at $\ts 2.10^{-4}$ by mass, is little 
different from that predicted by Big Bang nucleosynthesis. This is a major problem 
(Hata et al. 1995, Olive et al. 1995): either the Big Bang value is too high, or the
evolution of low-mass stars is wrong. 
\par In this paper we identify a mechanism by which low-mass stars destroy
(on the FGB) the \het\  that they initially produce during their MS evolution. 
Amusingly, this mechanism is driven by the \het\ itself, in a narrow zone just 
above the main hydrogen-burning shell that is characteristic of FGB stars.

\section{A Local Molecular-Weight Inversion}
\par Once the SCZ has reached its deepest extent, part-way up from the
base of the FGB, it retreats, and can be expected to leave behind a region
of uniform composition with the \het\ abundance of $\ts 1.8.10^{-3}$ as seen
in Fig. 1b.
This region is stable to convection according to the usual Schwarzschild 
criterion, and is quite extensive in radius although small in mass. The 
H-burning front moves outwards into the stable region, but preceding the 
H-burning region proper is a narrow region, usually thought unimportant, in 
which the \het\ burns. The reaction that mainly consumes it is 
$$  ^3{\rm He}\ (^3{\rm He}, 2{\rm p})^4{\rm He}\ ,\eqno(1)$$
which is an unusual reaction in stellar terms because it {\it lowers} the mean 
molecular weight: two nuclei become three nuclei, and the mean mass per 
nucleus decreases from 3 to 2. The molecular weight being the mean mass per
nucleus, but including also the much larger abundances of \ho\ and \hef\ that
are already there and not taking part in this reaction, this leads to an
inversion in the molecular-weight gradient. The effect is tiny (see Fig. 2a, 
from the same 1D simulation as Fig. 1): it is in about the fourth decimal place. 
But our 3D modeling shows it to be hydrodynamically unstable, as we should 
expect from the classic Rayleigh-Taylor instability.

\vskip 0.2truein
\psfig{figure=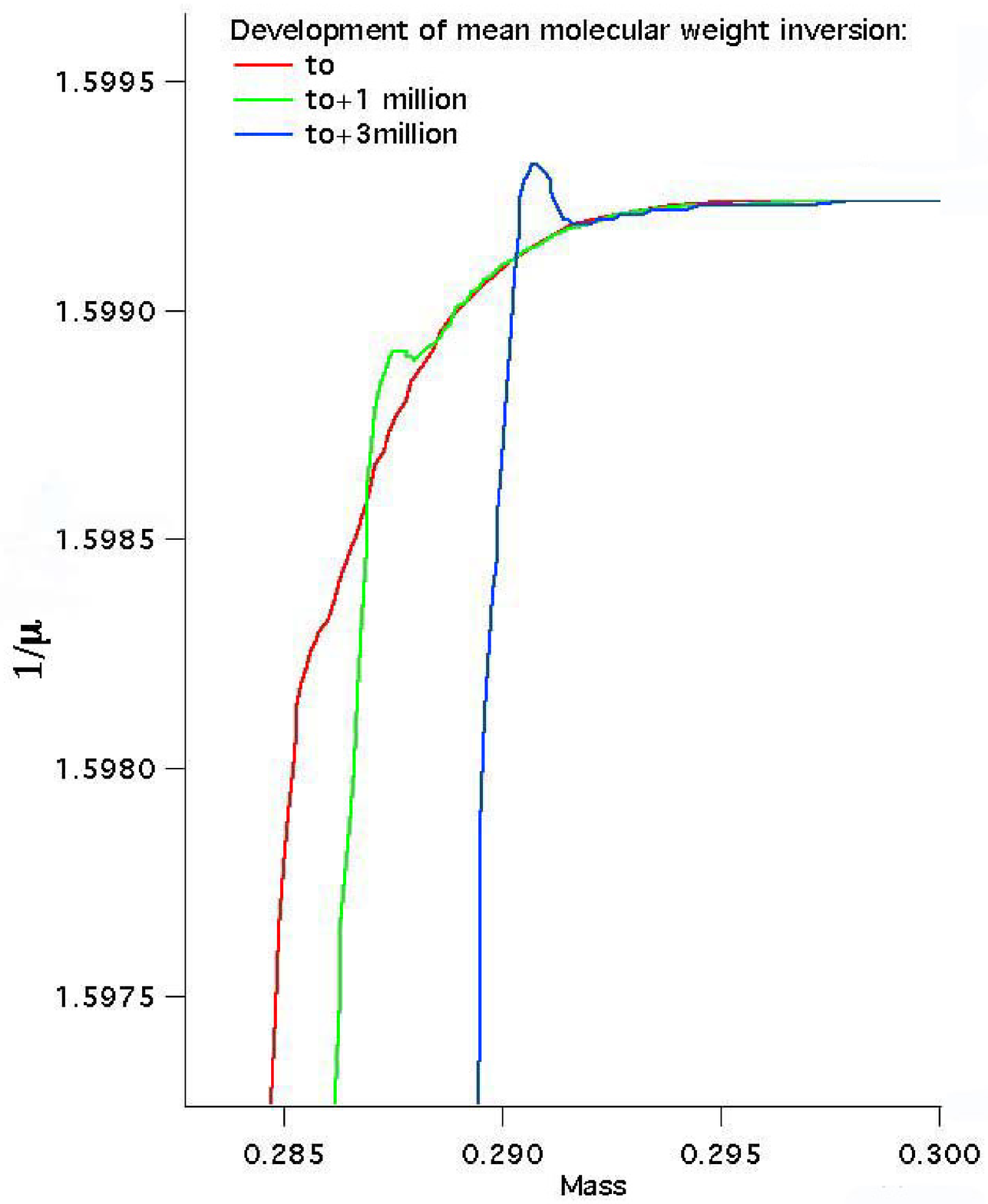,height=3.0in,bbllx=0pt,bblly=0pt,bburx=750pt,bbury=770pt,clip=}
\vskip -3.0truein
\hskip 2.7truein
\psfig{figure=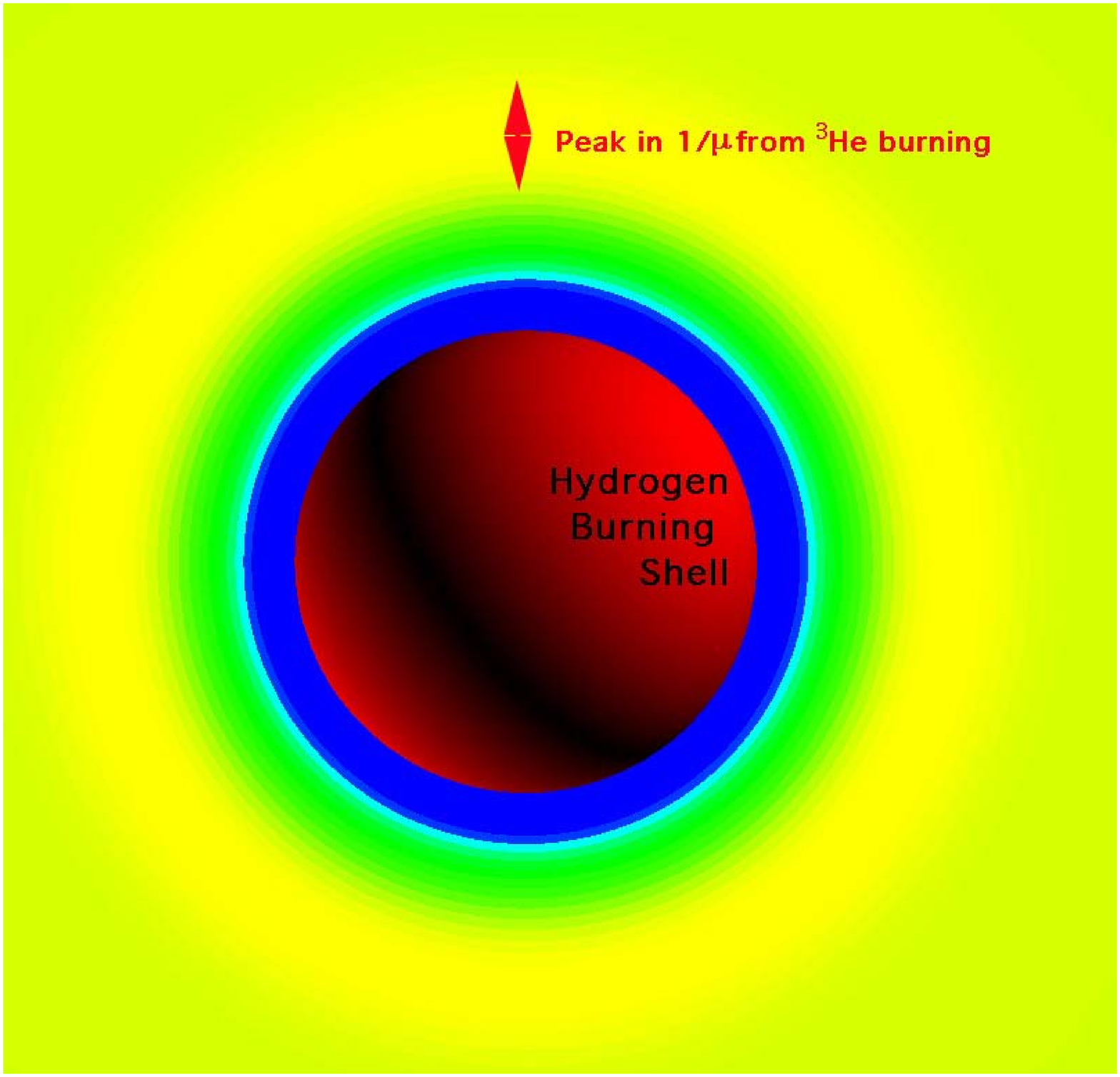,height=2.7in,bbllx=0pt,bblly=0pt,bburx=750pt,bbury=800pt,clip=}
\vskip 0.4truein
{\small {\bf Figure 2}. (a) The profile of (reciprocal) molecular weight, on a greatly
exaggerated vertical scale, as a function of mass coordinate. Once the burning 
shell has burnt outwards
to the region where the major composition has been homogenized by the
previous action of the SCZ, a small peak due to \het-burning, just outside
the main burning region, begins to stand out. The peak exists because
\het-burning {\it decreases} the mean molecular weight of the interacting
particles. (b) A cross-section through the central part of the star at the 
beginning of the 3D run. Mean molecular weight is color-coded, except that
the main hydrogen-burning shell is replaced by a red sphere. The 
$\mu$-inversion is the marked yellow band.} 
\vskip 0.1truein

\par Fig. 2a shows the $1/\mu$ profile plotted against mass coordinate. At a
relatively early stage (red curve) there is no bump, but just a slight
distortion at about 0.286$\Msun$. This is because the \het\ consunption is
taking place in a region where there is still a substantial $\mu$ gradient
left over from earlier history. But as the H-burning shell moves out (in mass), the
\het-burning shell preceding it moves into a region of more uniform \ho/\hef\ 
ratio, and so the peak in $1/\mu$ begins to stand out. By the time the shell
has moved to $0.289\Msun$ there is a clear local maximum in $1/\mu$, which 
persists indefinitely as the H-burning shell advances and the convective 
envelope retreats.
\par At this point in the evolution of our 1D star we mapped it on to a
3D model and used the hydrodynamic code `Djehuty' developed at the Lawrence 
Livermore National Laboratory (Baz{\'a}n et al. 2003, Eggleton et al. 2003, 
Dearborn et al. 2006). This code is described most fully in
the third of these papers. Although Djehuty is designed to deal with an entire
star, from center to photosphere, we economised on meshpoints by considering
only the region below the SCZ. The actual model selected was a Pop I star
of $1\Msun$ rather than a Pop II star of $0.8\Msun$, but there is very little
difference in regard to the \het\ behavior, and the peak in $1/\mu$.
\par Fig 2b is a color-coded plot of $\mu$ on a cross-section through the
initial 3D model. The shell where
the $\mu$-inversion occurs is the yellow-orange region sandwiched
between a pale green and a rather darker green. The inversion is at a radius of
$\ts 5.10^7\thn$m. The base of the SCZ is at $\ts 2.10^9\thn$m, well outside
the frame, and the surface of the star is at $\ts 2.10^{10}\thn$m. Notwithstanding
the red sphere shown in Fig 2b, the entire interior of the the star below the
SCZ was in the computational domain.
\section{Rayleigh-Taylor Instability}
\par Fig 3 shows the early development of the initially-spherical shell on
which $1/\mu$ has a constant value near its peak. After only $\ts 800\thn$secs,
the surface has begun to dimple, and by $2118\thn$secs the dimpling is very
marked, and the surface has begun to tear. Some points have moved $\ts 2\%$
radially, ie $\ts 2.10^5\thn$m, indicating velocities of $\ts 100\thn$m/s. 
The mean velocity decreases slightly in the passage from the second to the 
fourth panel. Other spherical shells, well away from the inversion on either 
side, show no such dimpling, at least until the influence of the inversion has 
spread to them.
\par The motion appears turbulent, and has the effect of diluting the
inverse molecular-weight gradient, but it cannot eliminate it. As the
turbulent region entrains more of the normally stable region outside it yet
below the normal convective envelope, it brings
in fresh \het, which  burns at the base of this mixing region, thus sustaining
the inverse molecular-weight gradient. Ultimately this turbulent region will
extend to unite with the normally-convective envelope, 
so that the considerable reservoir of \het\ there will also be depleted.
If its speed of $\ts 100\thn$m/s is maintained the time for processed material 
to reach the classically unstable SCZ is only about 5 months, while the time to 
burn through the  $\ts 0.02\Msun$ layer is over $10^6\thn$yrs.
\par Normal convective mixing, as in the usual SCZ, is a rapid
process: it would homogenise the surface layers in a matter of weeks, if it
were not already homogeneous by this stage in the star's evolution. Our new
mixing process might be expected to be very much slower, but in order to 
modify progressively the
composition in the normal convective envelope it need only operate on roughly 
the nuclear timescale of the star, which is $\ts 200\thn$Megayear at this point. 
\par However, small as is the $\mu$ inversion that drives our extra mixing,
it produces velocities that are surprisingly large, and in fact comparable
to the velocity of the normal convection. This is for two reasons. Firstly, 
although we might expect the inversion to be diluted to a trivial amount by 
the mixing that it produces, it is in fact sustained because the fresh \het\ that
is brought in by the mixing is burnt quite rapidly by reaction (1) inside
the inversion. If $\dm$ is the height of the peak, if $\tmix$ is the
timescale of mixing due to the turbulent motion, and if $\tburn$ is the
timescale of \het-burning in the inversion, then we expect that
$\dm$ is diluted by a factor $\ts\tmix/\tburn$. We estimate this below, finding
values of $\ts 10^{-4}$. Since $\dm\ts 10^{-3}$, the strength of our `engine' 
is roughly $\ts 10^{-7}$. Secondly, the normal convection of the outer envelope
is driven by what is in fact only a very small excess of actual temperature
gradient over the adiabatic gradient. The fractional excess of temperature
gradient is itself $\ts 10^{-5}$ near the base of the SCZ, and is 
therefore not, as one might expect, a great
deal larger than the fractional deficit in $\mu$ that drives our extra
mixing. Thus it is reasonable to expect our extra mixing to produce
turbulent velocities of perhaps a tenth of the normal convective mixing.

\vskip 0.1truein
\centerline{\psfig{figure=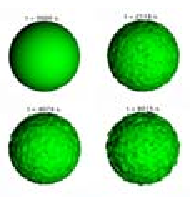,height=4.5in,bbllx=0pt,bblly=0pt,bburx=60pt,bbury=57pt,clip=}} 
\vskip -0.2truein
{\small {\bf Figure 3.} The development with time of a contour surface of mean 
molecular weight near the peak in Fig. 3. The contour dimples, and begins
to break up, on a timescale of only $\ts 2000\thn$sec. } 
\vskip0.1truein

\par As a first approximation, we can expect the effect of buoyancy due to
$\dm$ to generate a velocity of order
$$v^2\ts gl\thn{\dm\over\mu}\ ,\eqno(2)$$
where $g$ is the local gravity and $l$ is the distance an eddy moves before
dissolving into the ambient material; $l$ is normally estimated as the pressure
scale height. In ordinary convection, near the base of 
the SCZ, the equivalent estimate is
$$v^2\ts gl\thn{\dt\over T}\ ,\eqno(3)$$
where $\dt$ is the excess temperature, relative to the ambient medium,
of an eddy that expands adiabatically on rising. Near the base of 
the SCZ we have $g\ts 10^2\thn$m/s$^2$, $l\ts 10^{\thn 8.5}\thn$m, 
$\dt/T\ts 10^{-5}$, so that $v\ts 10^{\thn 2.8}\thn$m/s. In our first 
approximation (2), we ought to modify $\dm$ by the factor $\tmix/\tburn$ 
indicated above, and since we can estimate $\tmix$ as $l/v$, our second 
approximation is
$$v^3\ts gl^2\thn{\dm\over\tburn}\ .\eqno(4)$$
We find $g\ts 10^{\thn 4.5}\thn$m/s$^2$, $l\ts 10^7\thn$m, 
$\tburn\ts 10^8\thn$s and so $v\ts 10^{\thn 2.5}\thn$m/s. 
Thus the expected velocity driven by the inverted $\mu$-gradient is barely a 
modest factor of 2 down on the normal convective velocity near the base 
of the SCZ. This is smaller than the factor of $\ts 10$ in our first estimate 
mainly because $gl$ is an order of magnitude larger near the burning zone than
near the base of the SCZ.
\par The above argument establishes that the mixing is extended below the
classical Schwarz- schild limit, and that it is very fast compared to the 
nuclear
timescales of either the hydrogen-burning shell or the \het-burning reaction.
If the mixing is fast we can make an estimate of the total amount of \het\ 
that will be destroyed, by approximating the mixing region as homogeneous.
Beginning with a model near the location where the mechanism starts, and well
below the helium flash (top left in Fig. 1) the lifetime of the \het\ in a
region whose mass coordinate extends from $m_1$ to $m_2$ is
$${1\over t_{33}}= {1\over m_2-m_1}\thn\int_{m_1}^{m_2}\thn{dm\over\tau_{33}},
\eqno(5)$$
where 
$${1\over \tau_{33}} = {d \ln ^3{\rm He}\over dt}=\rho N(^3{\rm He})
\tl \sigma v\tg/2,\eqno(6)$$
$N(^3{\rm He})$ is the fractional local \het\ abundance, and 
$\tl\sigma v\tg$ is the 
thermally-averaged nuclear cross-section times velocity, i.e. reaction rate.
Similarly the timescale for the predominantly CNO-cycling hydrogen-burning
shell is 
$${1\over \tau_{\rm shell}}={N(^1{\rm H})\epsilon_{\rm CNO}(m_2-m_1)\over L}\ ,
\eqno (7)$$
where $N(^1$H) is the fractional hydrogen abundance above the shell and
$\epsilon$ is the local rate of nuclear energy generation. The result of such
integrations over some different mass regions shows that for any region much 
larger than the \het-burning region the ratio of lifetimes of \het\ destruction
and core-mass growth is constant and approximately 16. Thus for rapid mixing
the \het\ will be destroyed in 16 times as much mass as the hydrogen shell
burns through.

\par We believe that the extra mixing that we have discovered gives a
satisfactory answer to the problem mentioned in the second-last paragraph 
of the Introduction that
confronts Big Bang nucleosynthesis. Although low-mass stars do indeed
produce considerable amounts of \het\ on the MS, this will all be
destroyed by the substantially deeper mixing that we now expect on the FGB.
It is somewhat ironic that this deeper mixing is driven by the \het\ itself.
\par Our deeper mixing can also be relevant to further problems that have
troubled stellar modelers for several years. According to the classical models
of FGB stars, there is no further modification to the composition in an FGB
convective envelope after it has reached its maximum extent early on the FGB. 
Yet observations persistently suggest that the ratios $^{13}$C/$^{12}$C and 
$^{14}$N/$^{12}$C both increase appreciably as one goes up the FGB 
(Suntzeff 1993, Kraft 1994). Both
these ratios can be expected to increase only if the material in the envelope
is somehow being processed near the H-burning shell. Our model makes this very
likely. Although the $\mu$-inversion that we find is somewhat above the main
part of the H-burning shell, it is not far above and we can expect some modest
processing of $^{12}$C to $^{13}$C and $^{14}$N. According to Weiss \& 
Charbonnel (2004), it appears to be necessary for some extra mixing to take
place beyond the point on the FGB where the SCZ has penetrated most deeply; 
that is exactly the point where our mechanism should start to operate.
\par Correlations between abundance excesses and deficits of various elements
and isotopes in the low-mass evolved stars of globular clusters have been 
discussed thoroughly in (Kraft 1994). The
subject is complex, and it is hard to distinguish star-to-star variations
that may be due to evolution from those that may be due to primordial
variation. Evidence exists for both kinds of variation. However, we expect our
mechanism to lead to substantial evolutionary variations.

\section{Discussion}
\par The $\mu$-inversion that we investigate has not been noted before, to
our knowledge; but even if it has been noted previously 
it has probably been ignored because it is  small, and
because traditional 1D models only give turbulent mixing if they are
instructed to. We feel that our investigation demonstrates particularly
clearly the virtue of attempting to model in 3D, where the motion evolved
naturally, and to a magnitude that initially surprised us. 3D modeling is an 
expensive exercise, but we believe that it is amply justified. The mixing
process that we have identified appears to be capable of solving one
cosmological problem and two or more stellar problems.

\begin{acknowledgments}
We are grateful to R. Palasek for managing the code, and for help with
the graphics. This study has been carried out under the auspices of the U.S. 
Department of  Energy, National Nuclear Security Administration, by 
the University of  California, Lawrence Livermore National Laboratory, 
under contract  No. W-7405-Eng-48.
\end{acknowledgments}

\begin{discussion} 
\discuss{Weiss} {A comment:
The connection between the carbon anomalies and the $^3$He-problem
was made by Charbonnel (1995) and Weiss, Wagenhuber, and Denissenkov (1995)
coming from the empirical evidence for deep mixing.}

\discuss{Eggleton} {Yes: we refer explicitly to you paper with Charbonnel 
(2004). But you discussed mainly the possibility that the extra
mixing might be driven by differential rotation. Our point is that,
while rotation and differential rotation might influence the mixing,
perhaps accounting for some spread, a different mechanism (the Rayleigh-
Taylor mechanism which we describe) is bound to happen, and can account
for the $^3$He problem as well, perhaps, as for $^{13}$C and $^{14}$N.}

\discuss{Roxburgh} {In order to have the large $^3$He build up on the main
sequence you have to assume that this build up remains in place and is
not destroyed by the $^3$He epsilon-mechanism instability first proposed
by D. Gough (Nature, 240, 262, 1972).}

\discuss{Eggleton} {I believe that Gough's `solar spoon' mechanism was 
concerned with
the much deeper interior. Our mechanism kicks in relatively nearer the
surface.}
\end{discussion}

\end{document}